\documentclass[11pt,a4paper]{article}

\usepackage[utf8]{inputenc}
\usepackage[T1]{fontenc}
\usepackage{lmodern} 
\usepackage{graphicx} 
\usepackage{amsmath, amssymb, amsfonts} 
\usepackage{authblk} 
\usepackage{hyperref} 
\usepackage{cite} 
\usepackage{geometry} 
\usepackage{siunitx} 
\title{Hyper-spectral Imaging with Up-Converted Mid-Infrared Single-Photons}

\author[1]{Yijian Meng}
\author[1]{Asbjørn Arvad Jørgensen}
\author[1]{Andreas Næsby Rasmussen}
\author[2]{Lasse Høgstedt}
\author[2]{Søren M. M. Friis}
\author[1,*]{Mikael Lassen}
\affil[1]{Dansk Fundamental Metrologi A/S, Kogle Alle 5, 2970 Hørsholm, Denmark}
\affil[2]{NLIR Aps, Hirsemarken 1, 3520 Farum, Denmark}

\date{\today} 

\begin{document}

\maketitle

\begin{abstract}
Hyperspectral imaging in the mid-infrared (MIR) spectral range provides unique molecular specificity by probing fundamental vibrational modes of molecular bonds, making it highly valuable for biomedical and biochemical applications. However, conventional MIR imaging techniques often rely on high-intensity illumination that can induce photodamage in sensitive biological tissues. Single-photon MIR imaging offers a label-free, non-invasive alternative, yet its adoption is hindered by the lack of efficient, room-temperature MIR single-photon detectors. We present a single-photon hyperspectral imaging platform that combines cavity-enhanced spontaneous parametric down-conversion (SPDC) with nonlinear frequency up-conversion. This approach enables MIR spectral imaging using cost-effective, visible-wavelength silicon single-photon avalanche diodes (Si-SPADs), supporting room-temperature, low-noise, and high-efficiency operation. Time-correlated photon pairs generated via SPDC suppress classical intensity noise, enabling near shot-noise-limited hyperspectral imaging. We demonstrate chemically specific single-photon imaging across the \SIrange{2.9}{3.6}{\micro\meter} range on biological (egg yolk, yeast) and polymeric (polystyrene, polyethylene) samples. The system delivers high-contrast, label-free imaging at ultralow photon flux, overcoming key limitations of current MIR technologies. This platform paves the way toward scalable, quantum-enabled MIR imaging for applications in molecular diagnostics, environmental sensing, and biomedical research.
\end{abstract}

\section*{Introduction}
The development of label-free bioimaging techniques with high chemical specificity and minimal perturbation is important for advancing both fundamental biomedical research and clinical diagnostics \cite{diem2013molecular,shaked2023label, ghosh2023viewing}. Mid-infrared (MIR) light that covers the molecular fingerprint region from approximately 2.5 to 20 $\mu$m is uniquely positioned to address this need due to their ability to directly probe fundamental vibrational modes of molecular bonds \cite{shaw1999vibrational,zhao2012hydrogen, shi2020mid,pilling2016fundamental}. These intrinsic vibrational signatures provide unambiguous chemical contrast for key biomolecular components such as proteins, lipids, and nucleic acids, allowing detailed analysis of tissue composition, microbial presence, metabolic states, and disease-related biochemical changes \cite{carter2009vibrational,hackshaw2020vibrational}. Methods such as near-infrared Raman spectroscopy and Fourier transform infrared (FTIR) spectroscopy have demonstrated considerable potential \cite{nicolson2021spatially,fahelelbom2022recent,thomsen2022accurate}. However, biological tissues can be highly sensitive to optical exposure, and conventional high-intensity illumination can cause photodamage, induce molecular alterations, or disrupt physiological functions \cite{magidson2013circumventing, waldchen2015light}. To address these limitations, single-photon MIR imaging techniques have emerged as a promising alternative, providing label-free chemical contrast while substantially minimizing optical perturbation. This advancement offers a viable pathway to truly non-invasive molecular imaging in biological systems.

Traditional single-photon MIR imaging faces several critical limitations that hinder their broader adoption and performance \cite{dello2022advances}. A major challenge lies in the low quantum efficiency of MIR detectors, which often require cryogenic cooling to suppress dark counts and achieve acceptable sensitivity \cite{razeghi2014advances,wang2019sensing}. Common MIR detectors, such as mercury cadmium telluride or InSb photodiodes, exhibit significantly lower performance compared to their visible and near-infrared counterparts, especially at the single-photon level. More recently, emerging technologies such as superconducting nanowire single-photon detectors (SNSPDs) and transition-edge sensors (TESs) have demonstrated ultra-high sensitivity and low-noise performance in the MIR regime. However, they face similar drawbacks that require operation at cryogenic temperatures \cite{dello2022advances}. 

To overcome these challenges, different optical techniques have been explored to enable room temperature MIR bioimaging at the single-photon level. Such as undetected MIR photons using non-linear interferometers \cite{lemos2014quantum,kalashnikov2016infrared,kviatkovsky2020microscopy} and nonlinear up-conversion detection \cite{dam2012room,huang2022wide,ge2024quantum,junaid2019video}. When operated at the quantum level applying advanced detection strategies and quantum-correlated light sources, these approaches potentially facilitate high-contrast, noninvasive imaging with high sensitivity and precision. In the nonlinear interferometer approach, the spatial information of the sample is encoded in the reduction of interference contrast between photons generated in two consecutive nonlinear processes.
In contrast, up-conversion detection schemes convert MIR photons to the visible range via second-order nonlinear optical interaction, enabling the use of silicon single-photon avalanche diodes (Si-SPADs). This approach offers a high-efficiency, room-temperature alternative with compatibility with mature photonic technologies. Furthermore, by implementing spontaneous parametric down-conversion (SPDC) and coincidence detection between the signal and idler photons, the technique enables background noise suppression and single-photon sensitivity, making it particularly advantageous for low-light imaging and spectroscopic applications \cite{mancinelli2017mid}. Recent advances in this approach have demonstrated its potential for high-speed imaging and molecular fingerprinting, further expanding the applicability of MIR bioimaging in biomedical research \cite{zhao2023high,huang2022wide,junaid2019video}. 
Despite their distinct advantages, these two techniques present unique challenges. Microscopy with undetected photons requires precise interferometric alignment and suffers from limited photon conversion efficiency, while up-conversion detection relies on highly efficient nonlinear processes and precise timing synchronization. 

In this work, we present a novel MIR single-photon imaging platform for hyperspectral analysis of polymeric and biological samples. Using a cavity-enhanced SPDC source, we generate time-correlated photon pairs with tunable idler wavelengths from 2.9 to 3.6\,\textmu m. This tunability enables chemically specific hyperspectral imaging with single-photon sensitivity, allowing detailed molecular analysis while minimizing photodamage~\cite{pilling2016fundamental,hermes2018mid}. The signal (near-infrared) and idler (mid-infrared) photons are independently upconverted via second-order nonlinear processes to visible and near-infrared wavelengths and detected using Si-SPADs. Coincidence gating between channels effectively suppresses uncorrelated background, enhancing sensitivity at ultralow photon flux. This dual-path upconversion scheme provides efficient, room-temperature, low-noise detection of MIR single photons. Compared to typical MIR upconversion methods which rely on classical MIR sources and intensity detection~\cite{dam2012room,junaid2019video,zhao2023high}, our approach achieves high sensitivity, spectral resolution, tunability, and noise rejection—making it suitable for label-free hyperspectral imaging of biological samples under minimally invasive conditions.

\section*{Experimental configuration}

\begin{figure}[th]
    \centering
\includegraphics[width=1\textwidth]{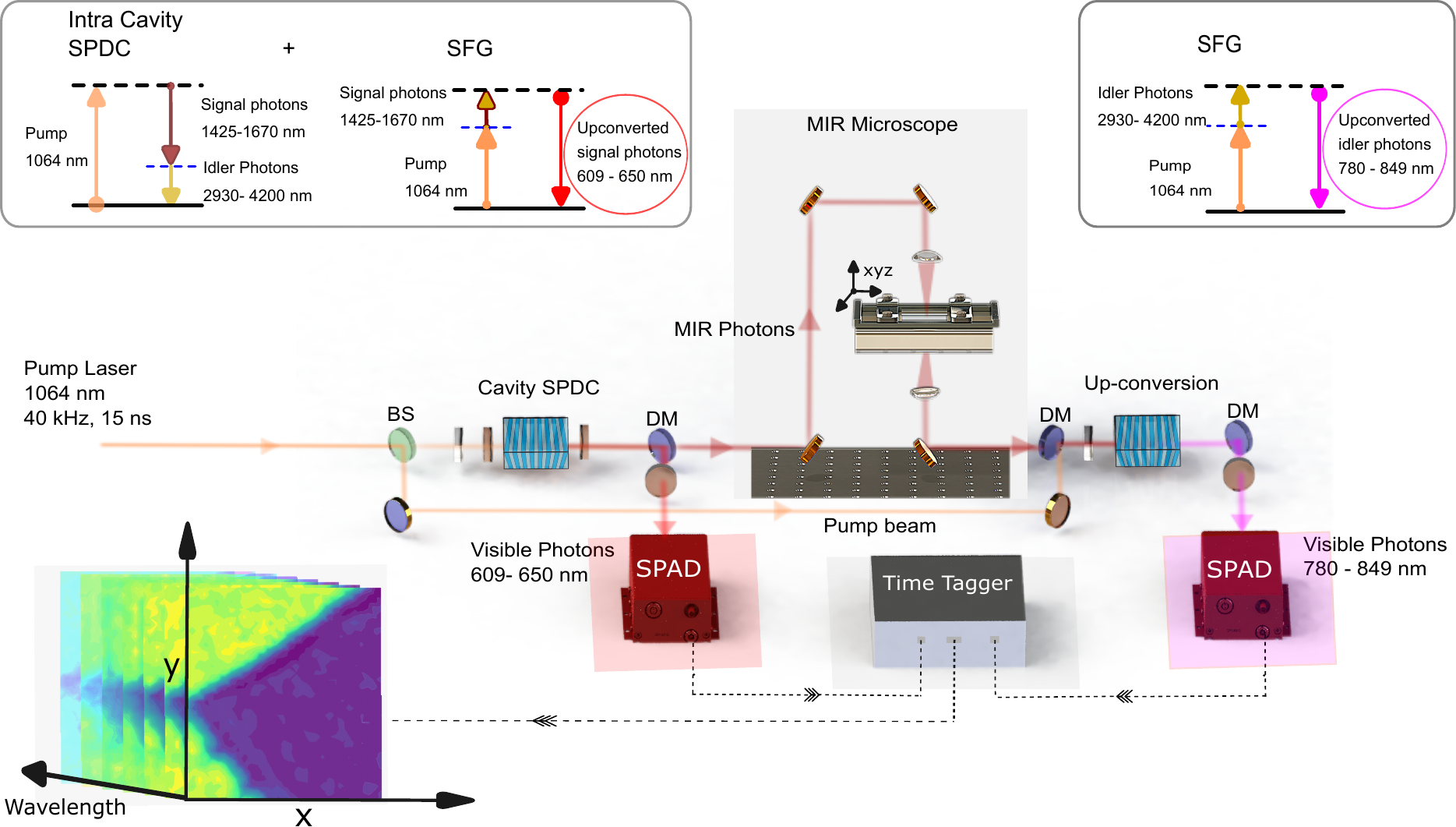} 
    \caption{Experimental setup for MIR single-photon hyperspectral imaging based on time-correlated single-photon counting (TCSPC) and cascaded nonlinear optical processes. A pulsed 1064 nm laser pumps two fan-out periodically poled lithium niobate (PPLN) crystals. The first crystal is embedded within a linear optical cavity to enable cavity-enhanced spontaneous parametric down-conversion (SPDC), producing time-correlated photon pairs comprising a mid-infrared (MIR) idler photon and a near-infrared (NIR) signal photon. The NIR photon is upconverted intracavity to the visible range via sum-frequency generation (SFG) and detected by a Si-SPAD. The MIR idler photon is directed onto a biological sample mounted on a motorized XYZ translation stage. The transmitted MIR photon is subsequently upconverted to the visible via a second SFG process in the second PPLN crystal and detected by a second Si-SPAD. Arrival times of both photons are recorded using a time tagger. Insets: energy-level diagrams illustrating the cascaded nonlinear interactions involved in intra-cavity SPDC and the sum-frequency upconversion of the signal and idler photons, respectively. The detected upconverted generated wavelengths are 609-650 nm for the signal photons and 780-849 nm for the idler photons in the full spectral scanning range of the SPDC. 
 }
    \label{fig:1} 
\end{figure}

The experimental setup is depicted in Fig.~\ref{fig:1} and consists of four main components: a tunable mid-infrared (MIR) single-photon source based on cavity-enhanced SPDC, a single-pass upconversion unit, high-efficiency Si-SPADs (Count, Laser Components) with quantum efficiencies exceeding 65\% for the upconverted photons, and a custom-designed MIR microscope with low-loss optics for MIR light. The photon-pair source is driven by an actively Q-switched nanosecond Nd:YAG laser (Sol compact Q-switched DPSS laser, Bright Solutions) operating at a central wavelength of \SI{1064}{\nano\meter} , with a pulse duration of 15 ns and a repetition rate of 40 kHz. The pump pulses are focused into a 40~mm long fan-out structured periodically poled lithium niobate (PPLN) crystal (HC Photonics), which is positioned at the center of a 55~mm long linear optical cavity. This cavity is formed by two identical dielectric mirrors: high-reflectivity (HR) coated for the NIR signal photons and high-transmission (HT) coated for both the idler photons and pump laser. Each mirror has a radius of curvature of 100~mm, resulting in a stable cavity configuration that produces a fundamental Gaussian mode with a beam waist of approximately 120~\textmu m at the center of the crystal. This configuration ensures optimal spatial mode overlap between the interacting waves and efficient nonlinear conversion within the PPLN crystal. The cavity is operated between 10 to 30\% of its self-oscillation threshold to ensure stable and low-noise generation of photon pairs via SPDC. The cavity SPDC is single-resonant for NIR signal photons (1400-1700 nm). Spectral tuning of the generated photon pairs is achieved by altering the phase-matching condition through translation of the fan-out structured PPLN crystal along its poling gradient using a high-precision stepper motor. This enables fast continuous wavelength tuning of the MIR idler photon in the 2.9–3.60 $\mu$m range. This wavelength range is particularly significant as it encompasses key ro-vibrational absorption bands corresponding to the fundamental stretching modes of biologically relevant functional groups, including O–H, C–H, and N–H bonds.  As a result, this MIR region is exceptionally well suited for label-free, chemically selective imaging of biological samples, which we will investigate in the following.


\subsection*{Cascaded nonlinear optical processes}

In Fig.~\ref{fig:1}, the insets depict the energy-level diagrams of the cascaded nonlinear optical processes central to our MIR imaging platform. The first stage involves the intra-cavity generation of signal and idler photons via SPDC, where a pump photon at wavelength $\lambda_p=1064$ nm is converted into a pair of lower-energy photons—a signal photon $\lambda_s=1510-1683$ nm and an idler photon $\lambda_i=2900-3600$ nm. The SPDC nonlinear interaction not only conserves energy but also establishes strong time and frequency correlations between the signal and idler photons, potentially enabling quantum-enhanced measurements. Within the same nonlinear crystal, the signal photons remain confined in the cavity and undergo a second nonlinear process—sum-frequency generation (SFG) with the \SI{1064}{\nano\meter} pump field. This cascaded interaction results in upconverted signal photons at wavelengths $\lambda_{\text{up,s}} = (\lambda_s\lambda_{\text{pump}})/(\lambda_s + \lambda_{\text{pump}})$, corresponding to wavelengths in the visible range (609–650 nm). These upconverted signal photons exit the cavity and are detected using Si-SPADs. The SFG process achieves an internal conversion efficiency of approximately 3–5\%, depending on the pump power. In parallel, the idler photons generated via intra-cavity SPDC exit the cavity and are directed to the microscope to interrogate the sample. After transmission, the idler photons are coupled into a second fan-out structured periodically poled lithium niobate (PPLN) crystal. To perform up-conversion, the idler photons are combined with a separate pump field at \SI{1064}{\nano\meter} using a dichroic mirror. The two beams are overlapped and co-propagated in a single-pass through the PPLN crystal, where SFG/upconversion occurs, producing upconverted photons with wavelength $\lambda_{\text{up,i}}$ in the near-infrared range (780–849 nm). The upconversion achieves a conversion efficiency of approximately 10-30\%, which depends on the pump power.These photons are subsequently detected using a second Si-SPAD. The phase-matching (movement) conditions of the upconversion PPLN crystal are synchronized with those of the SPDC crystal using calibrated stepper motors, ensuring spectral alignment across the imaging range. To suppress background noise and prevent contamination from stray light, a series of high-performance optical filters is employed prior to Si-SPADs. These filters are carefully selected to match the upconverted photon bandwidths and ensure efficient rejection of pump leakage, residual fluorescence, and ambient light, thereby enhancing the signal-to-noise ratio in single-photon detection. The arrival times of both the signal and idler photons are recorded using  a Time Tagger (Swabian Instruments), a time-correlated single-photon counting (TCSPC) module that enables high-resolution temporal gating and correlation measurements.



\section*{Results}
In the following, we present the characterization results of the MIR photon-pair source, along with demonstrations of low-photon-flux imaging based on time-correlated single-photon detection. The system enables rapid and continuous tuning of the MIR idler photon wavelength across the 2.9–3.6~$\mu$m range, enabling hyperspectral analysis of vibrational modes associated with different molecular functional groups. By implementing photon gating and correlation detection, we effectively reduce uncorrelated background noise while suppressing excessive noise of the pump laser. 


\subsection*{Source characterization and data Acquisition }

\begin{figure}[th]
    \centering
\includegraphics[width=0.8\textwidth]{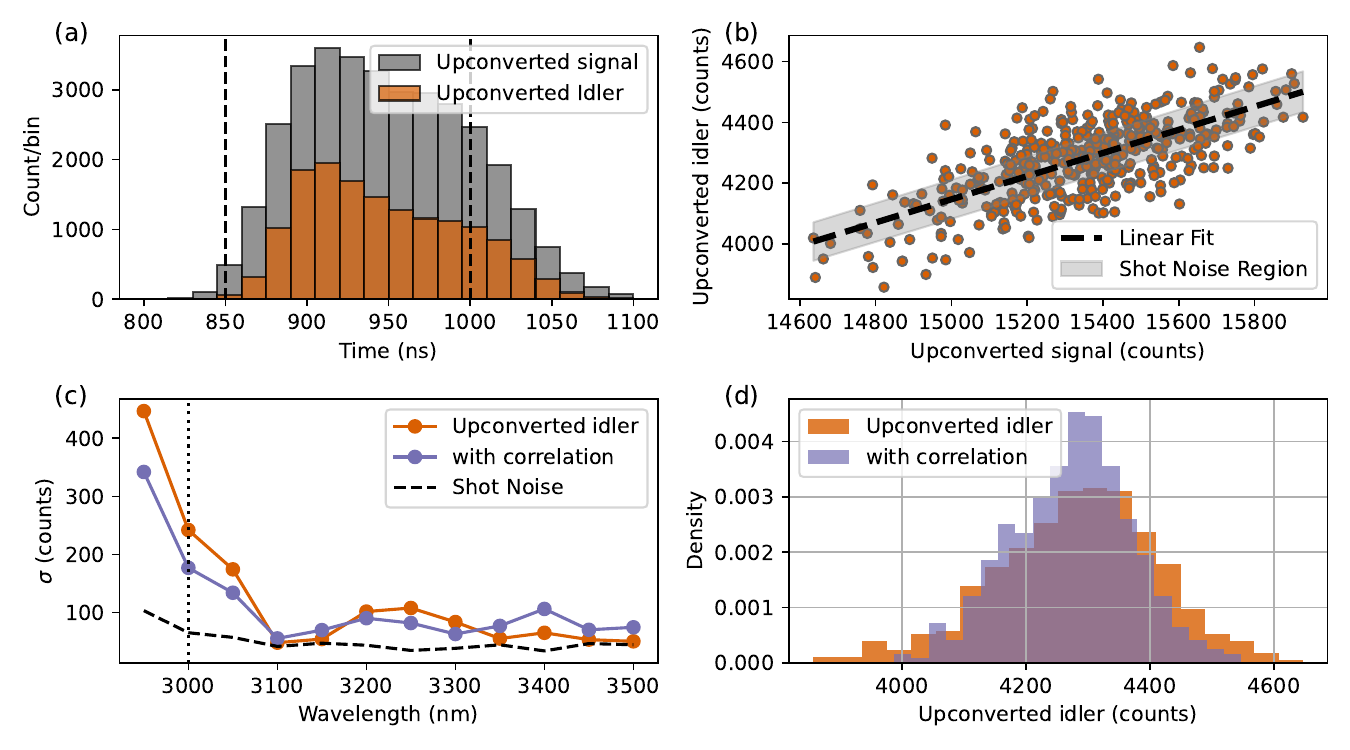} 
    \caption{Characterization of noise reduction and stability enhancement via signal-idler correlation. \textbf{(a)} shows the raw histograms of photon arrival times from the signal and idler detectors at \SI{2950}{\nano\meter}. A \SI{150}{\nano\second} gating window (black dashed lines) is applied to both channels to suppress background noise. Arrival times are recorded on a time tagger relative to a trigger signal derived from the pulsed laser. \textbf{(b)} Correlation between detected counts in the signal and idler arms at \SI{3000}{\nano\meter} , with a linear fit (black dashed line) \textbf{(c)} Standard deviation of idler counts before (orange) and after (purple) normalization using signal-idler correlation. Improvement is observed at wavelengths where fluctuations significantly exceed the shot noise level (black dashed line). \textbf{(d)} Histogram of idler counts before and after normalization at \SI{3000}{\nano\meter}. The full width at half maximum (FWHM) is reduced after applying correlation-based rescaling, indicating enhanced stability.}
    \label{fig2:example} 
\end{figure}

A representative measurement of the SPDC source with upconverted photons is shown in Fig.~\ref{fig2:example}a. A \SI{150}{\nano\second} detection gate is applied in each experimental cycle to suppress background noise and improve detection sensitivity. Figure~\ref{fig2:example}b) shows a strong intensity correlation between the gated photon counts at the two detectors. Photon counts at both detectors are integrated for a duration of \SI{2}{\second}. The shaded region indicates the shot noise level corresponding to the upconverted idler count. Fig~\ref{fig2:example}c compares the intensity noise of the gated photon counts (orange) with the shot noise limit (black dashed line). In spectral regions where excess intensity noise dominates, the correlation between the two detectors can be exploited to suppress this noise. This suppression is evident below \SI{3100}{\nano\meter}, as indicated by the purple line in Fig.~\ref{fig2:example}c. The rescaled upconverted idler count is computed as:
\begin{equation}
    N_{up,i}^{\text{corr}} = N_{up,i}^c \cdot \frac{\langle N_{up,s}^c \rangle}{N_{up,s}^c},
\end{equation}
where $N_{up,s}^c$ and $N_{up,i}^c$ are the gated counts of the upconverted signal and idler photons, respectively, and $\langle N_{up,s}^c \rangle$ is the average upconverted signal count.
The effectiveness of this noise mitigation strategy depends on the trade-off between the additional shot noise introduced by the upconverted signal and the excess intensity noise present in the upconverted idler counts. An example of this rescaling applied to data at \SI{3000}{\nano\meter}, where excess noise is particularly significant, is shown in Fig.~\ref{fig2:example}d. These measurements demonstrate that correlated photon detection combined with a rescaling strategy suppresses excess intensity noise in the upconverted idler channel and highlights the potential of correlation-based noise mitigation to improve sensitivity in upconversion-assisted SPDC imaging. 

\subsection*{Spectral calibration}

\begin{figure}[h]
    \centering
    \includegraphics[width=1\textwidth]{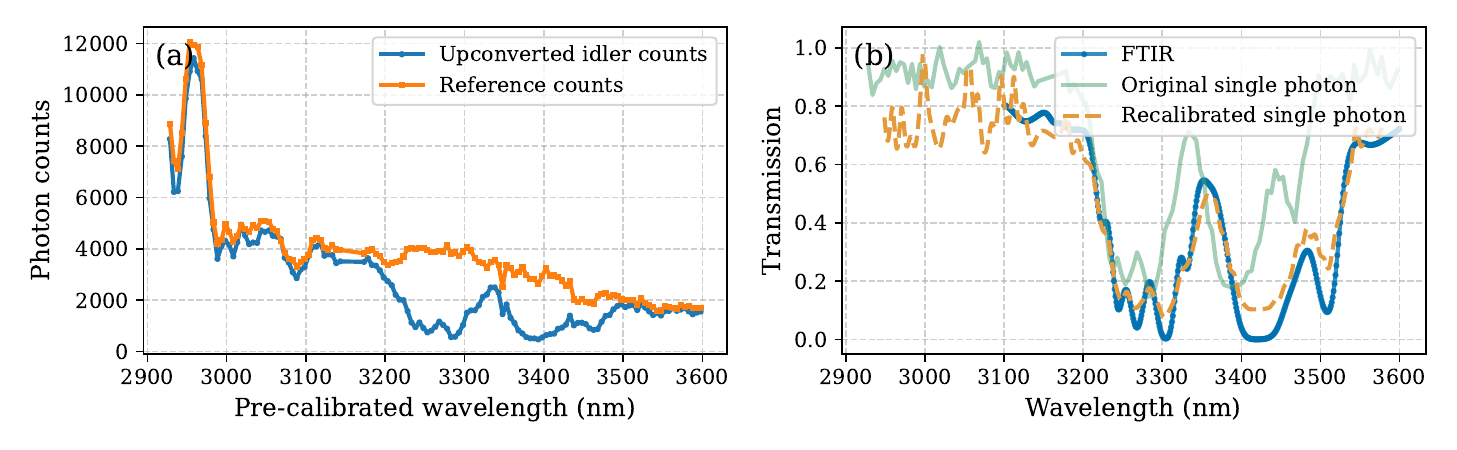} 
        \caption{Normalizing and calibrating the spectroscopy. \textbf{a} Comparison of detected photons counts with and without a polystyrene sheet. The counts without the sample are measured during imaging to account for wavelength-dependent up-conversion efficiency. Transmission is calculated as the ratio of these counts. \textbf{b} Calibration of the MIR wavelength axis was performed using FTIR spectral data from a reference polystyrene sheet. The solid black trace represents the reference FTIR spectrum, acquired under controlled conditions to ensure accurate spectral positioning of characteristic vibrational modes. The solid green trace corresponds to the raw MIR single-photon spectrum obtained using our imaging system, prior to any spectral correction. The Spectra are measured using a wavelength step size of \SI{7}{\nano\meter}. The dashed trace shows the MIR single-photon spectrum after wavelength calibration, aligned to match the FTIR reference. This calibration procedure ensures accurate spectral assignment and consistency across subsequent hyperspectral imaging measurements.}
    \label{fig:Calibration} 
\end{figure}

The MIR spectra obtained from various sample types exhibited distinct absorption peaks corresponding to O--H and C--H molecular bonds. First the measured spectroscopic traces were normalized by comparing the detected photon counts with and without a polystyrene sheet, as shown in Fig.\ref{fig:Calibration}a. The counts without the sample were recorded during imaging to account for wavelength-dependent upconversion efficiency and background spontaneous noise. Transmission was then calculated as the ratio of the sample counts to the reference counts. Further, the initial wavelength is determined based on the mean spectrum of the up-converted signal light field, obtained through Gaussian fitting, while the pump light operates above threshold. The idler wavelength is then extrapolated based on the pump laser wavelength, measured on a traceable wavelength meter (HP 86120B). For MIR imaging, the pump laser is tuned below threshold intensity, and the crystal's motor position is used to correlate with the wavelength determined in the previous step. To ensure more accurate spectral calibration, we employed a polystyrene (PS) standard reference sample. The PS sample was measured using both the single-photon imaging system and a commercial Fourier Transform Infrared (FTIR) spectrometer (Thermo Scientific, Nicolet iS50 FTIR) using a spectral resolution of \SI{8}{\micro\meter}$^{-1}$, see Fig.\ref{fig:Calibration}b. The spectrum obtained from the single-photon imaging setup, measured with a \SI{7}{\nano\meter} step size and comprising 100 data points, was aligned with the FTIR reference spectrum by calibrating the wavelength axis using a fitting function derived from the Beer–Lambert law:

\begin{align}
\widetilde{T}_C (\lambda',Rx)&= 1 - \left( 1 - \widetilde{T}_{\text{SP}} (\lambda',x) \right)^R, \\
\widetilde{T}_{\text{SP}} (\lambda,x) &= 1 - e^{-\alpha(\lambda) x}
\end{align}
where $\widetilde{T}_C$ ($\widetilde{T}_{\text{SP}}$) denotes the normalized transmission function with (without) rescaled wavelength $\lambda'=(a\lambda+b)$, where $a$ ($b$) represents the first- and zero-order corrections, respectively; $x$ denotes the thickness of the sample and $\alpha$ denotes the absorption coefficient as a function of wavelength. The calibration function also includes $R$,  which accounts for the exponential dependence of absorption on the varying sample thickness $x$ across different measurements.  We fit $\widetilde{T}_C (\lambda',Rx)$ to the FTIR reference spectrum using three free parameters: $a$,$b$, and $R$.  This function serves as the baseline spectrum against and will be used for calibrating all acquired single-photon hyperspectral images. Fig.~\ref{fig:Calibration}b illustrates an example of the fitting procedure, with the grey line indicating the raw normalized absorption spectrum and the dashed orange line representing the spectrum after rescaling to match the FTIR reference (blue line).  To ensure consistency across measurements and minimize systematic error, the same set of wavelength correction was applied uniformly to all acquired spectra and images during the calibration process. 

\subsection*{Hyperspectral imaging of Polymer samples}

\begin{figure}[h]
    \centering
    \includegraphics[width=1\textwidth]{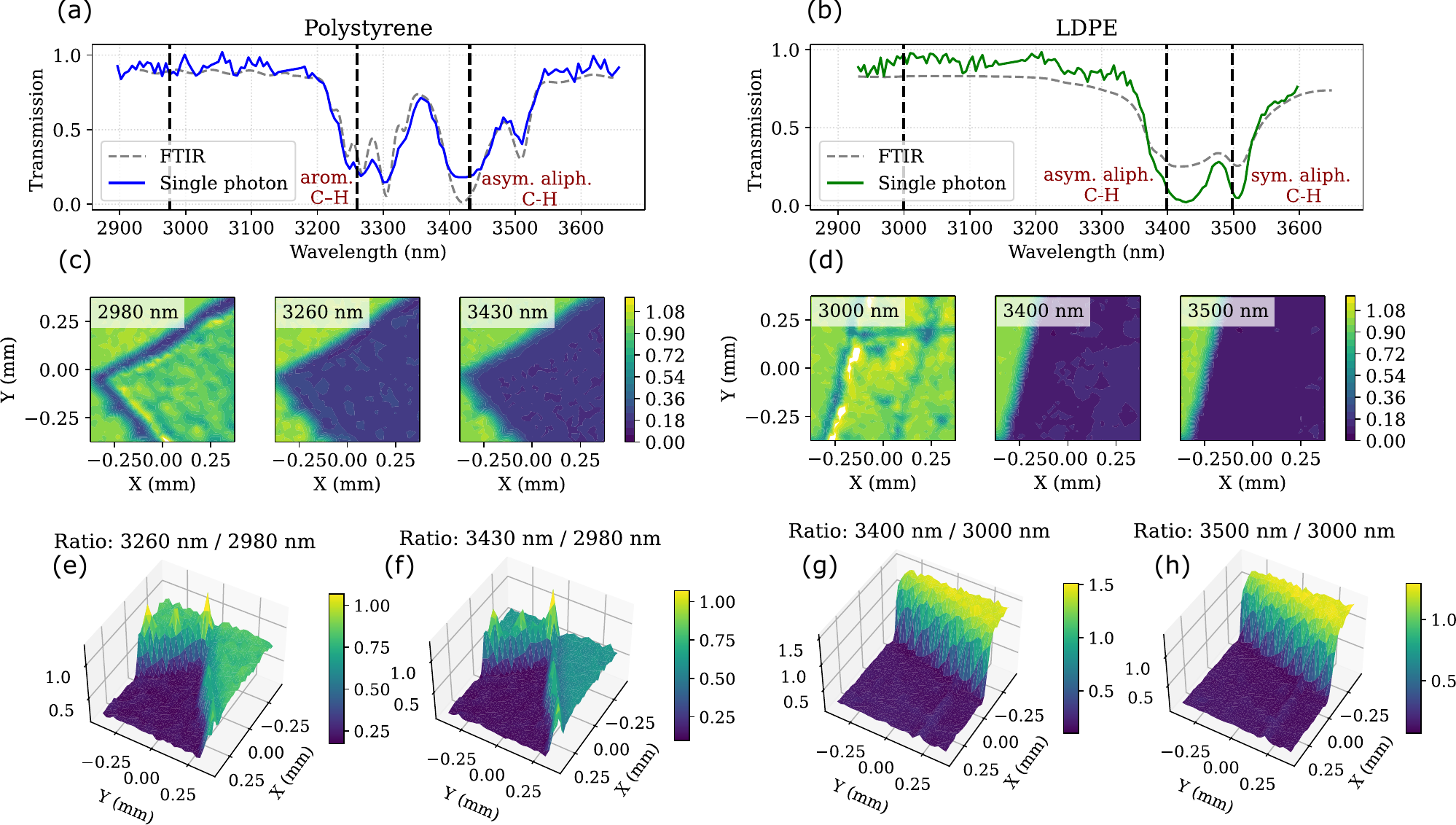} 
        \caption{Mid-infrared (MIR) absorption analysis of polymers: \textbf{a} Polystyrene sheet and \textbf{b} low-density polyethylene (LDPE). In both figures, the dashed curved lines represent rescaled FTIR spectroscopy results for comparison. The vertical dashed lines in \textbf{a} and \textbf{b} mark the wavelengths corresponding to the three spatial images shown in \textbf{c} and \textbf{d}, respectively. The contrast image between selected wavelengths is presented in \textbf{e-h}, demonstrating good sensitivity to the relevant C--H vibrational transitions.Each spatial image has an area of \SI{775}{\mu\meter} by \SI{775}{\mu\meter} with \SI{25}{\mu\meter} step size.}
    \label{fig:PolymerSample} 
\end{figure}

The results of the spectroscopic and hyperspectral imaging of the polymer samples are presented in Fig.\ref{fig:PolymerSample}. For each sample, we measure a series of spatial image between \SI{2900}{\nano\meter} and \SI{3600}{\nano\meter} with a \SI{50}{\nano\meter} wavelength spacing, thus 20 hyper-spectral plans are measured. Each spatial image has an area of \SI{775}{\micro\meter} by \SI{775}{\micro\meter} with a step size of \SI{25}{\micro\meter}, which is approximately the size of the MIR beam focused on the sample.  Each measurement point takes \SI{2}{\second}.  Due to the wavelength-varying phase-matching condition at the non-linear crystal,  we scan outside of the sample region to obtain a reference photon level, mitigating the nonuniform photon conversion efficiency across different wavelengths. The reference photon scan occurs before and after each spatial scan to mitigate, to the first order, the drift of photon level during imaging. The MIR optical power used for these measurements was approximately 2 fW, corresponding to an average photon flux of 40k photons per second, or roughly one MIR photon per pulse. Fig.~\ref{fig:PolymerSample}a shows transmission dips in aromatic C-H stretches and asymmetric aliphatic C-H stretches in polystyerene, while Fig.~\ref{fig:PolymerSample}b shows asymmetric aliphatic C-H stretches and symmetric aliphatic C-H stretches in low density polyethylene (LDPE). The transmission spectra in Fig.~\ref{fig:PolymerSample}a,b are measured with a $\approx$\SI{8}{cm^{-1}} spectral resolution. The curved dashed line indicates the fitted spectra obtained from FTIR, with one free parameter that accounts for the sample thickness variation. The vertical dashed lines in Fig.~\ref{fig:PolymerSample}a,b indicate the selected spectral positions for spatial imaging shown in Fig.~\ref{fig:PolymerSample}c,d, respectively. The high-contrast images from Fig.~\ref{fig:PolymerSample}e-h indicate excellent material-specific absorptions in both polymer samples. These results demonstrate the capability of our MIR single-photon imaging system to resolve distinct vibrational absorption features in polymer samples with high spectral and spatial resolution. The strong agreement with the FTIR-fitted spectra and the high-contrast images at selected wavelengths confirm material-specific absorption signatures in both polystyrene and LDPE, highlighting the potential of the system for label-free chemical imaging.

\subsection*{Hyperspectral imaging of egg yolk and yeast cells}

\begin{figure}[h]
    \centering
    \includegraphics[width=1\textwidth]{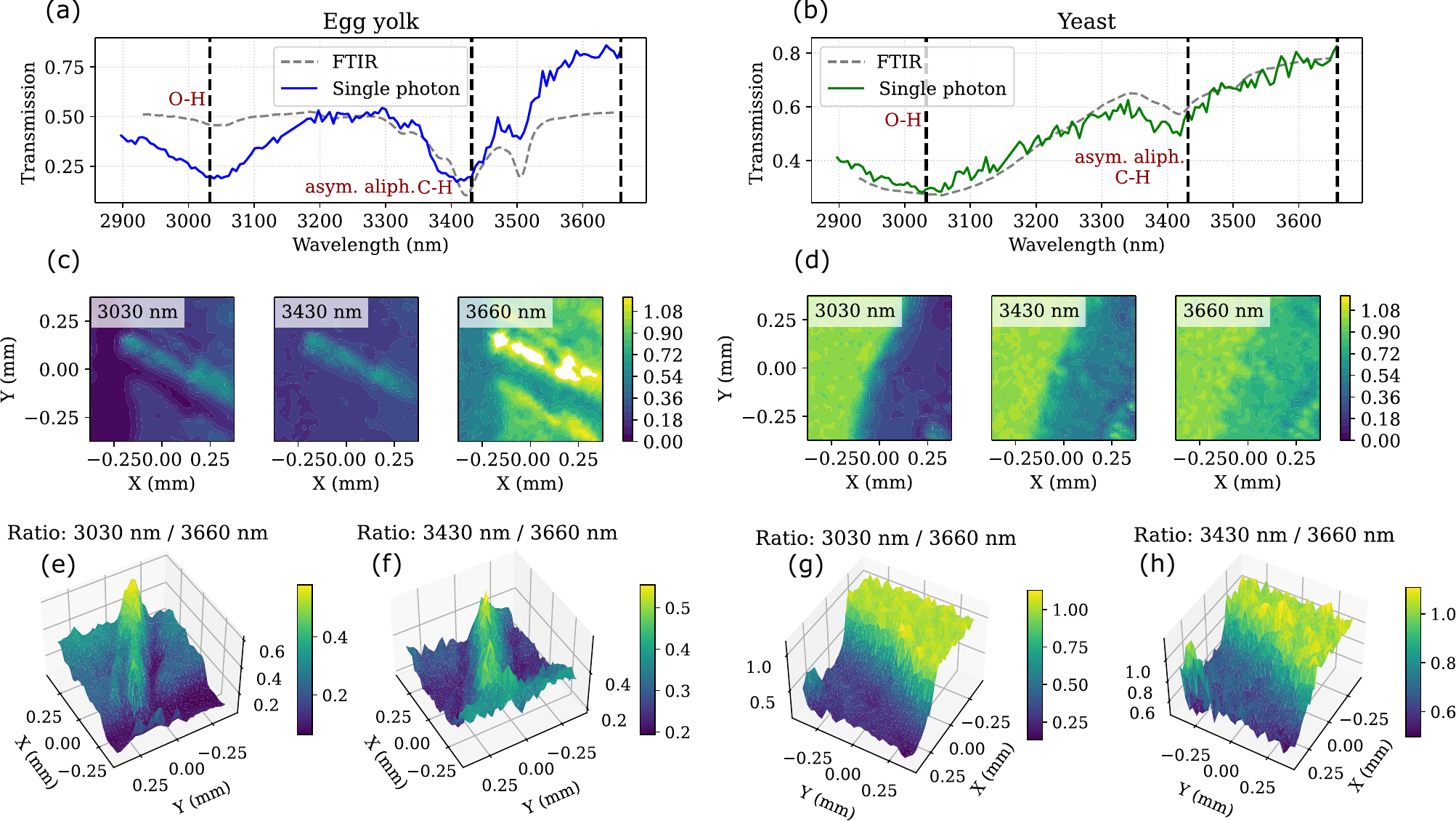} 
        \caption{MIR transmission analysis of biological samples: \textbf{a} Egg yolk and \textbf{b} yeast. The dashed curved lines show rescaled FTIR spectra for reference, with vertical dashed lines marking the wavelengths of the spatial images in \textbf{c} and \textbf{d}. Contrast images in \textbf{e-h} highlight sensitivity to O--H and C--H vibrational transitions. Each spatial image has an area of \SI{775}{\micro\meter} by \SI{775}{\micro\meter} with \SI{25}{\mu\meter} step size.
        }
    \label{fig:BioSample} 
\end{figure}

In addition, measurements of egg yolk and yeast samples are presented in Fig.\ref{fig:BioSample}. Fig.~\ref{fig:BioSample}a shows transmission dips corresponding to O–H stretching (3030 nm) and asymmetric aliphatic C–H stretching (3430 nm) vibrations in egg yolk, while Fig.~\ref{fig:BioSample}b displays similar features for the yeast sample. The curved dashed lines represent FTIR reference spectra, which show good agreement with the single-photon spectra. To accommodate the semi-solid nature of biological samples, the FTIR measurements were performed in attenuated total reflectance(ATR) mode, which may account for the relatively shallow O–H dip observed in the egg yolk spectrum. The contrast images in Fig.~\ref{fig:BioSample}e–h, derived from the spatial maps in Fig.~\ref{fig:BioSample}c,d at the indicated spectral positions (vertical dashed lines in Fig.~\ref{fig:BioSample}a,b) clearly reveal material-specific absorption features at key vibrational wavelengths. The single-photon spectra show good agreement with FTIR reference data, and the contrast images highlight material-specific absorption at key wavelengths.

\section*{Discussion and outlook}
Single-photon imaging in the MIR region has shown significant promise for label-free biological imaging. This technique operates at sub-picowatt photon flux levels, enabling non-invasive spectral and spatial information from biological tissues and polymer samples without the need for exogenous labeling agents. The demonstration of hyperspectral imaging over the 2.9 -\SI{3.6}{\micro\meter} range on biological samples such as yeast cells and egg yolk samples, as well as polymer samples including polystyrene and polyethylene, highlights the chemical specificity achievable with this platform. The strong agreement with FTIR spectra confirms the fidelity of our measurements, while single-photon sensitivity ensures minimal sample perturbation, an essential feature for applications involving sensitive biological materials or dynamic processes. A key strength of our approach lies in the integration of cavity-enhanced SPDC with nonlinear frequency up-conversion, facilitating the use of standard off-the-shelf available Si-SPADs. This room-temperature operation combined with high quantum efficiency and low dark counts overcomes the primary limitations of conventional MIR detectors, which typically require cryogenic cooling and exhibit inferior performance at the single-photon level. The time-correlated photon pairs generated by SPDC enable coincidence gating, effectively suppressing classical background noise and enhancing the signal-to-noise ratio to near the shot-noise limit. With its effective noise suppression and high sample-specific sensitivity at ultralow flux, we foresee this approach extending beyond material characterization, offering potential in non-invasive detection in areas such as biomedical diagnostics, environmental sensing, and chemical analysis. These attributes establish it as a versatile and powerful tool for advancing modern mid-IR photonics.

Future work will focus on enhancing imaging resolution, for example, by incorporating near-field scanning techniques that enable sub-diffraction-limited imaging of bacterial cells. Additionally, the current imaging speed is limited by the repetition rate of the pump laser (\SI{40}{kHz}); Employing a megahertz pump laser could yield an improvement of up to three orders of magnitude, allowing sub-minute hyperspectral imaging speeds, an essential advancement for high-throughput applications. Additionally, nonlinear up-conversion efficiency and phase-matching bandwidth impose spectral constraints and affect overall sensitivity. Continued optimization of non-linear crystal design, such as custom poling structures or novel materials, could broaden the spectral coverage and improve conversion efficiency. Advances in synchronization and timing electronics will further refine coincidence detection, reducing noise and improving temporal resolution. Coupling this imaging modality with advanced data analysis techniques, such as machine learning and artificial intelligence, promises to extract deeper chemical and structural insights from complex hyperspectral datasets. This integration could accelerate discovery in molecular biology, materials science, and clinical practice.

In summary, our results demonstrate label-free chemically selective imaging of polymers and biological samples at ultralow photon flux, confirming the system’s potential for noninvasive analysis of sensitive specimens. By overcoming key challenges in MIR single-photon detection and imaging, this approach opens a new avenue for ultrasensitive, chemically specific imaging and, with further improvements in speed, resolution, and spectral coverage, is poised to become a versatile tool in mid-infrared photonics.

\section*{Methods}

\subsection*{Sample Preparation}
Polymer samples of polystyrene (PS) and low-density polyethylene (LDPE) were cleaned and mounted on infrared-transparent substrates suitable for mid-infrared transmission measurements. Biological samples, including yeast and egg yolk, were prepared as thin semi-solid layers to ensure sufficient optical transmission in the MIR region. 

\subsection*{Data Acquisition and Scanning Procedure}
For each sample, hyperspectral spatial images were recorded by scanning the MIR wavelength in increments of approximately \SI{50}{\nano\meter} between \SI{2900}{\nano\meter} and \SI{3600}{\nano\meter}. At each wavelength, spatial scanning was performed on a \SI{775}{\micro\meter} by \SI{775}{\micro\meter} area using a motorized stage with \SI{25}{\micro\meter} step size in both lateral directions. Each spatial pixel was integrated for \SI{2}{\second} to accumulate sufficient photon counts for reliable imaging. Due to wavelength-dependent phase-matching conditions in the nonlinear crystals affecting photon conversion efficiency, reference scans were acquired outside the sample area before and after each spatial scan to record baseline photon levels. These reference measurements were used to normalize the spatial images, correcting for nonuniform conversion efficiency and slow drift in photon flux during acquisition.

\subsection*{Data Processing and Analysis}
Time gating between signal and upconverted idler photons was applied to suppress background counts and enhance signal-to-noise ratio. Transmission spectra at each spatial location were constructed by normalizing the detected photon counts at each scanned wavelength individually.
Spectral features were compared with reference FTIR spectra obtained from standard laboratory instruments. FTIR measurements on biological samples were performed in ATR mode to accommodate their semi-solid form. The spectral resolution of the hyperspectral images was approximately \SI{8}{\micro\meter}$^{-1}$, limited by the spectral bandwidth of the SPDC source and the up-conversion phase-matching bandwidth. Spatial contrast images were generated at selected spectral bands corresponding to key molecular vibrational modes identified in the transmission spectra. The contrast of the image and the chemical specificity were evaluated to demonstrate the capability of the platform for the differentiation of biological samples and material.

\subsection*{AI assisted copy editing} 
AI were used solely to enhance the clarity, readability, and stylistic consistency of the manuscript text, ensuring correctness in grammar, spelling, punctuation, and tone. All scientific content, interpretations, and conclusions remain the responsibility of the authors.

\bibliographystyle{unsrt} 
\bibliography{sample}

\clearpage

\section*{Acknowledgements}
This research was funded by the Danish Agency for Institutions and Educational Grants and QuRaman project under QuantEra supported by Innovation Fund Denmark (1116-00003B) and the European Partnership on Metrology (21GRD07 “PlasticTrace”). 

\section*{Author information} 
Danish Fundamental Metrology, Kogle Allé 5, 2970, Hørsholm, Denmark
Yijian Meng, Asbjørn Arvad Jørgensen, Andreas Næsby Rasmussen and Mikael Lassen
NLIR Aps, Hirsemarken 1, 3520 Farum, Denmark
Lasse Høgstedt and Søren M. M. Friis

\section*{Author contributions statement}
Y.M. and M.L. designed the experiment. Y.M. performed the measurements. Y.L. and M.L. analysed the data. Y.M., A.A.J. and A.N.R. developed the software for XYZ stage control and synchronization of the OPO and upconversion source. L.M. and S.M.M.F. developed the upconversion unit. Y.M. and M.L. wrote the initial manuscript with input from all authors. M.L. acquired funding, conceived the project and supervised the research. 

\section*{Additional information}

\section*{Competing interests.}
The authors declare no competing financial interests. 

\section*{Data Availability Statement}
The data that support the findings of this study are available from the corresponding author upon reasonable request.

\section*{Correspondence}
Correspondence and requests for materials should be addressed to Y.M. (yme@dfm.dk) and M.L. (ml@dfm.dk).

\end{document}